\documentclass[a4paper,12pt]{article}
\usepackage{times}
\usepackage{epsfig}
\usepackage{amsfonts}
\usepackage{amsmath}
\usepackage{cancel}
\usepackage{graphicx}
\usepackage{mathcomp}
\usepackage{placeins}
\usepackage{floatflt}
\usepackage{float}
\usepackage{amssymb}
\usepackage{subfigure}

\title{Mean extinction times in cyclic coevolutionary rock-paper-scissors dynamics}
\author{Markus Sch\"utt and Jens Christian Claussen\footnote{Corresponding author.}
\\
Institut f\"ur Neuro- und Bioinformatik
\\
 Universit\"at zu L\"ubeck, D-23562
L\"ubeck, Germany
}
\date{February 24, 2010}

\begin{document}

\maketitle
\abstract{
Dynamical mechanisms that can stabilize the coexistence or diversity in biology are generally of fundamental interest. 
In contrast to many two-strategy evolutionary games, games with three strategies and cyclic dominance like the 
rock-paper-scissors game (RPS) stabilize coexistence and thus preserve biodiversity in this system. 
In the limit of infinite populations, 
resembling the traditional picture of evolutionary game theory, 
replicator equations predict the existence of a fixed point in the interior of the phase space. 
But in finite populations, strategy frequencies will run out of the fixed point because of stochastic fluctuations, 
and strategies can even go extinct. For three different processes and for zero-sum and non-zero-sum 
RPS as well, we present results of extensive simulations for the mean extinction time
(MET), depending on the number of agents $N$, 
and we introduce two analytical approaches for the derivation of the MET.
}

\section{Introduction}
There are several examples of cyclic dominance in biology: Males of the californian lizards \textit{Uta stansburiana} are known to 
inherit
 three different mating strategies which cyclically dominate each other 
\cite{eidechsen1,eidechsen2}.
Another example is given by three strains of \textit{Escherichia coli} 
\cite{E. coli Raum 1}, in tropical marine ecosystems 
\cite{marine cyclic evolution} or in high arctic vertebrates 
\cite{vertebraten}. Cyclic dominance is also important in some theoretical models like the susceptible-infected-recovered-susceptible 
(SIRS) model in epidemiology 
\cite{SIRS1}, in cyclic extensions of the Lotka-Volterra model 
\cite{mckane05,butlergoldenfeld09}
or the Public Goods Game 
\cite{public goods,loners}. Here, cyclic dominance is a way to preserve the coexistence of strategies (what is often called `biodiversity' in a biological context) in the models.

A straightforward model system for cyclic dominance is the rock-paper-scissors game (RPS) well known from schoolyards. It contains the three strategies `rock', `paper', and `scissors' with a simple domination rule: Paper wraps rock, scissors cut paper, rock crushes scissors. The payoff a dominating strategy gets from the dominated one is set to $1$ for normalization, the payoff for a tie is $0$, and a dominated strategy gets a 
payoff $-s$, where we assume $-s<0$. Hence the payoff matrix of this game reads
\begin{equation}
P
=
\begin{pmatrix} 0 & -s & 1\\ 1 & 0 & -s \\ -s & 1& 0 \end{pmatrix}.
\end{equation}
For the standard choice $s=1$ we have a zero-sum RPS, for all other values the game is non zero-sum. One can simply intuit the impact of $s$ 
\cite{CT08}: For huge $s$ it is important for a player to avoid loosing, so it is
successful playing the same strategy as the majority;
hence an equilibrium that includes all three strategies is 
unstable. On the other hand, for small $s$ it is more important to win occasionally so that the mixed equilibrium becomes stable. Experiments indicate $s<1$ for the lizard example
\cite{Eidechsen kleiner 1} and $s>1$ for \textit{E. coli} 
\cite{E. coli Raum 1}.

For a long time, the resulting dynamics have been described by a meanfield approximation in the limit of infinite populations. The traditional way to describe such evolutionary games was the standard replicator equation, an equation of motion for the density of an arbitrary strategy $\alpha$ 
\cite{Replikatorgleichung},
\begin{equation}
\dot x_\alpha=\psi x_\alpha(\pi^\alpha(\vec{x})-\langle \pi(\vec{x})\rangle),
\end{equation}
where $\psi$ is a constant prefactor which can be absorbed in the time scale by a 
constant rescaling of time. $\pi^\alpha(\vec{x})=x_1 P_{\alpha,1}+x_2P_{\alpha,2}+x_3P_{\alpha,3}$ and $\langle\pi(\vec{x})\rangle$ are the payoff of strategy $\alpha$ and the average payoff of the whole population, respectively. The adjusted replicator equation 
\cite{adjusted replicator equation}
\begin{equation}
\dot x_\alpha=\frac{x_\alpha}{\Gamma+\langle\pi(\vec{x})\rangle}(\pi^\alpha(\vec{x})-\langle\pi(\vec{x})\rangle),
\end{equation}
has been used less frequently. The prefactor $1/(\Gamma+\langle\pi(\vec{x})\rangle)$ can be absorbed in the time scale by a 
dynamical rescaling of time for symmetric conflicts like the RPS,
preserving stability properties.
For asymmetric conflicts, like Dawkins Battle of the Sexes 
\cite{Battle of the Sexes}, this is not possible, 
as the average payoffs in each population do not coincide;
hence the adjusted replicator equation can lead to qualitatively modified dynamics 
\cite{TCH05,C07}.
The derivation of the replicator equations and 
Fokker-Planck-equations (comprising the first-order corrections)
for the intrinsic noise are commonly taken from the master equation of
the underlying discrete stochastic process 
\cite{helbing92,helbing93,TCH05,mckane05,chalubsouza}.

Both replicator equations predict the existence of a fixed point at $\left(x_R,x_P,x_S \right)=\left(\frac{1}{3},\frac{1}{3},\frac{1}{3}\right)$. The FP is neutrally stable for $s=1$, asymptotically stable for $s<1$, and 
unstable for $s>1$. In the case of finite populations of size $N$, the population can drive out of the fixed point because of stochastic fluctuations, and after some time one of the strategies will go extinct. Once this has happened, a second strategy will die out soon because of the lack of cyclic dominance, and the third strategy will therefore survive forever. Several efforts have been made to overbear this shortcoming especially of the zero-sum RPS, for example spatial discretization of populations (for a review see 
\cite{Graphenevolution 3}), mobility 
\cite{mobility}, the introduction of intelligent update rules (best response 
\cite{best response}), the introduction of the parameter $s$ as mentioned above 
\cite{CT08}, or the computation of a critical system size above which coexistence of strategies is likely 
\cite{TCH05}, but it is still an open question how long it takes on average until the first strategy has gone extinct. 
In this paper we investigate the scaling of the mean time to the extinction of the first strategy (mean extinction time, MET), depending on the system size $N$ and the parameter $s$, for well mixed populations and three evolutionary processes (with two of them having the standard and accordingly the adjusted replicator equation as limits $N\to\infty$), and present two analytical approaches which give theoretical insight in the scaling of the MET.

\section{Evolutionary processes}
Contrary to replicator equations describing the dynamics of relative abundance
densities, real populations are finite (and discrete),
appropriate modeling thus bases on discrete stochastic processes
(of birth and death).
In the classical Moran process 
\cite{original_Moran} an individual $a$ is chosen with probability proportional to its fitness. It reproduces, and the offspring replaces another randomly chosen individual $b$. In the frequency-dependent Moran process 
\cite{moran} which 
extends the classical Moran process \cite{original_Moran} by
considering coevolution, the fitness is not static but depends on the frequencies of the strategies. For better comparison with the processes mentioned 
beneath, in each time step we choose an individual $a$ at random which reproduces with probability
\begin{equation}
\phi_{M}^{\alpha(a)} (n_{\alpha},n_{\beta},n_{\gamma})=\frac{1}{2}\frac{1-\omega+\omega \pi_{\alpha(a)}}{1-\omega+\omega\left\langle \pi\right\rangle},
\end{equation}
and the offspring replaces another randomly chosen individual $b$. 
Greek lettes $\alpha,\beta,\gamma$ can assume each of the
strategies $R, S, P$, respectively.
Here, $\omega$ is the selection strength, the $n_{i}$ are the number of agents playing strategy $i$, $\pi_{\alpha}=(n_1P_{\alpha,1}+n_2P_{\alpha,2}+n_3P_{\alpha,3})/(N-1)$ the payoff for an individual of strategy $\alpha$, and $\langle \pi\rangle $ the average payoff of the whole population. In the limit $N\to\infty$, 
the Moran process leads to the so-called adjusted replicator equation 
\cite{TCH05}. Note that a factor $\frac{1}{2}$ is introduced here for a better comparability with the two processes mentioned beneath.

The Moran process is a well-established stochastic process for evolutionary
birth-death (for growing population sizes, see e.g.\ \cite{poncela09})
dynamics with overlapping generations 
and therefore 
(in its frequency-dependent extension) serves as a standard
model of evolutionary game theory.
However, the Moran process requires perfect global information 
about the whole population, an assumption that can be unrealistic and undesirable. 
Because of that, two local processes have been mentioned. Again we choose an individual $a$ at random for reproduction. Another randomly chosen individual $b$ changes its strategy to the strategy of $a$ with probability
\begin{equation}
\phi_{lu}^{\beta(b)\rightarrow \alpha(a)}(n_{\alpha},n_{\beta},n_{\gamma})=\frac{1}{2}+\frac{\omega (\pi_{\alpha(a)}-\pi_{\beta(b)})}{2\Delta\pi_{max}}
\end{equation}
for the local update 
\cite{TCH05} and
\begin{equation}
\phi_{F}^{\beta(b)\rightarrow \alpha(a)}(n_{\alpha},n_{\beta},n_{\gamma})=\frac{1}{1+e^{-\omega (\pi_{\alpha(a)}-\pi_{\beta(b)})}} 
\end{equation}
for the Fermi process 
\cite{fermi}, respectively, where $\Delta\pi_{max}$ is the maximum of the possible payoff difference. In the limit $N\to\infty$, these processes lead to the standard replicator equation 
(local update) 
\cite{TCH05}, 
and a nonlinear replicator equation (Fermi process) 
\cite{TPN07}
with similar properties,
respectively.
The common approach for the derivation of the equations of motion 
and the first-order corrections for demographic stochasticity are 
to derive a Fokker-Planck equation from the master equation
for the respective stochastic process
\cite{helbing92,helbing93,TCH05,C07,chalubsouza}.

Here we focus on the mean extinction time for the Moran and Local Update
processes.
For each process, the probability of increasing the population strength of the strategy $\alpha$ by $1$ in a single time step and decreasing the population strength of strategy $\beta$ by $1$, is given by
\begin{equation}
T_{process}(n_{\alpha},n_{\beta},n_{\gamma}\rightarrow n_{\alpha}+1,n_{\beta}-1,n_{\gamma})=\frac{n_{\alpha}}{N}\frac{n_{\beta}}{N}\phi_{process}^{\beta\rightarrow\alpha}(n_{\alpha},n_{\beta},n_{\gamma}).
\end{equation}
This quantity is known as hopping rate. Note that the sum over all hopping rates is $\neq 1$ because reactions are possible that do not lead to changes in strategy frequencies. 
The time scale is chosen so that one reaction takes place every unit time step, 
including empty steps with no strategy change.
\FloatBarrier

\section{Mean extinction times}

\begin{figure}
\subfigure[]{\epsfig{file=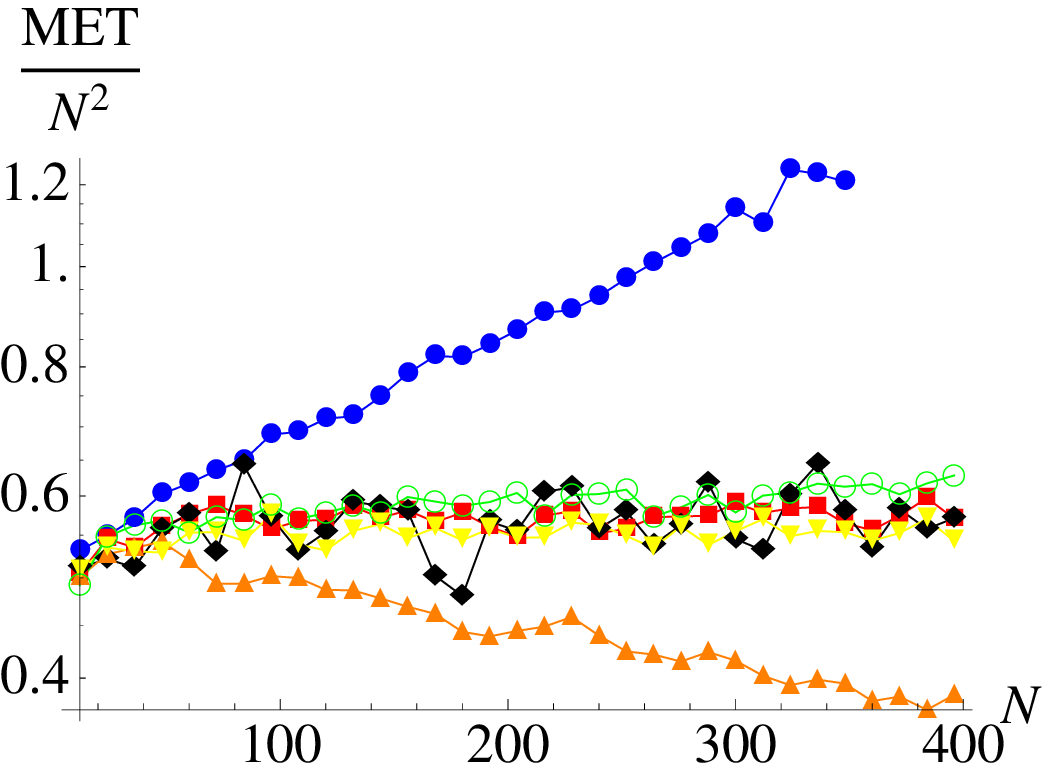,height=35mm}}
\subfigure[]{\epsfig{file=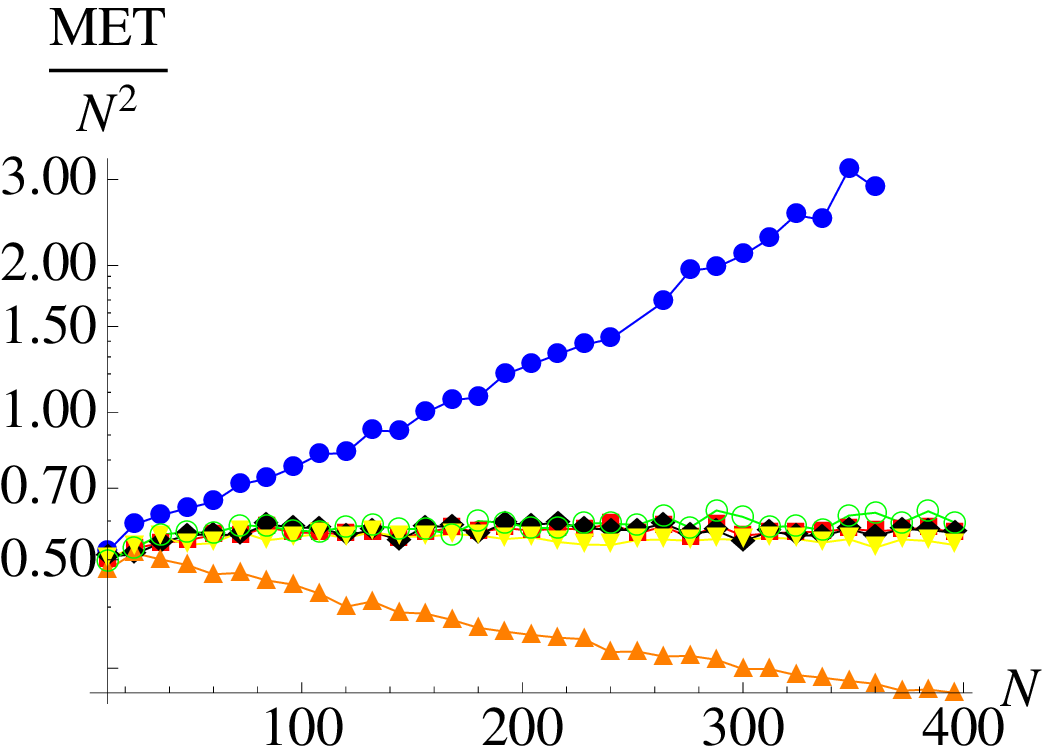,height=35mm}}
\subfigure[]{\epsfig{file=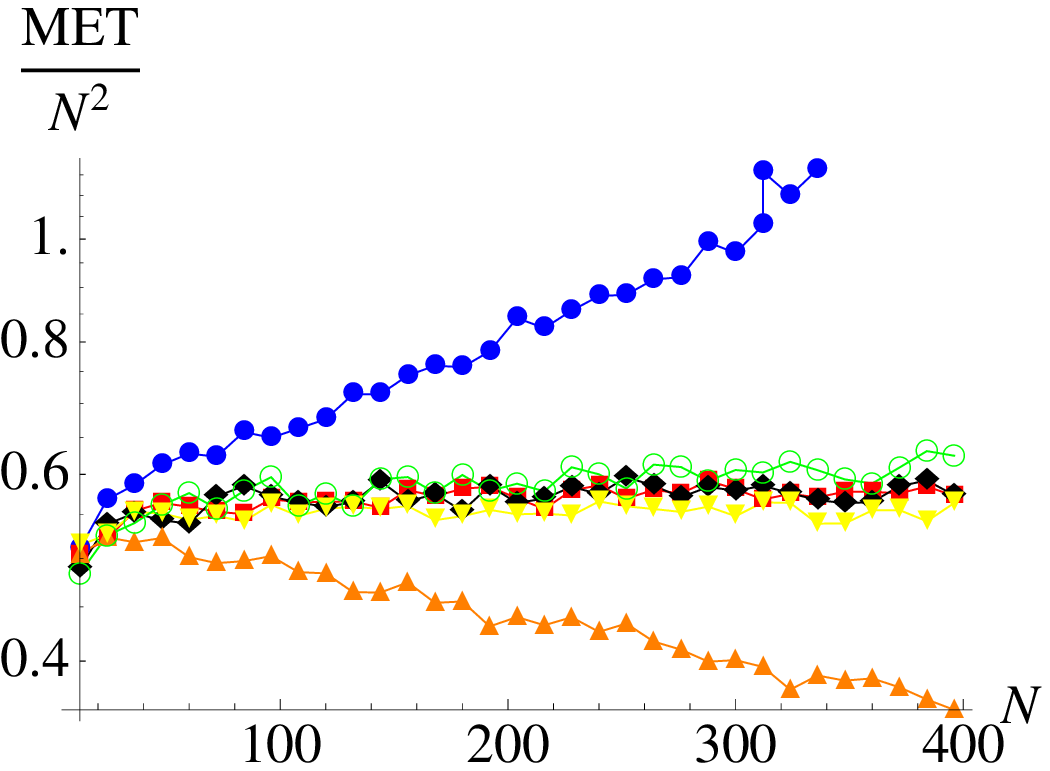,height=35mm}}
\caption{Semi-logarithmic plots of the simulation data for the METs for the local update process (a), the Moran process (b) and the Fermi process (c). The METs were divided by $N^2$. Each point is an average over the extinction times of 1000 simulations. Red squares: $s=1.0$, $\omega=0.50$, black rhombi: $s=1.0$, $\omega=0.05$, orange upwards triangles: $s=1.2$, $\omega=0.50$, yellow downwards triangles: $s=1.2, \omega=0.05$, green open circles: $s=0.8$, $\omega=0.05$, blue filled circles: $s=0.8$, $\omega=0.50$. For the Moran process $\omega=0.45$ has been used instead of $\omega=0.50$ to keep the transition 
probabilities in the interval $\left[0,1\right]$.}
\label{MET}
\end{figure}

Compared to the time scale a mutant occurs -- re-introducing a strategy -- 
the time scale characterizing the survival properties of a genetic strain is
defined by the mean extinction time.
For two strategies, and fixed $N$, one has a onedimensional Markov process
for which a closed expression allows to proceed analytically
as has been demonstrated 
by Antal and Scheuring \cite{antalscheuring}.
For higher-dimensional cases unfortunately exact general solutions
of the mean extinction time (or a mean-first-passage time \cite{redner})
problems are not known;
in this case the situation is furthermore 
hampered by the location-dependent dynamics within a simplex boundary.
So in most cases one has to rely on systematic numerical investigations.
We have carried out extensive simulations to quantify the mean times until one of the three strategies has gone extinct. 
We have analyzed the mean extinction times for all three described processes depending on the number of agents $N$ and the parameters $s$ and $\omega$. 
In general, we observe the following properties (see Fig.~\ref{MET}):
For all three processes the MET becomes independent of the selection strength $\omega$ for $s=1.0$. 
Further on, the MET for $s=1.0$ is identical for all three processes and proportional to $N^2$,
\begin{equation}
\langle t_{ext}\rangle(s=1)\approx (0.54\pm0.02) N^2.
\end{equation}

As expected, for $s<1$ the MET is larger and for $s>1$ smaller than for $s=1$, but for weak selection the difference is small. In a logarithmic plot of the MET divided by $N^2$ one can easily see that $\langle t_{ext}(s=1.0)\rangle\propto N^2$, but also $\langle t_{ext}(s<1)\rangle\propto N^2 e^{f(s,\omega)N}$ with $f(s,\omega)>0$ and independent of $N$. For the Moran process the stabilizing effect for $s<1$ is larger than for the other processes.

On the first view, for $s>1$ the MET seems to be $\langle t_{ext}(s>1)\rangle\propto N^2 e^{-f(s,\omega)N}$. But such a proportionality would imply a disappearance of the MET for large $N$. 
This is unrealistic because even the shortest way would require $N/3$ steps. For that a dependence $\langle t_{ext}(s>1)\rangle=c_1 N^2+c_2N^2 e^{- f(s,\omega)N}$ is more likely (with $f(s,\omega)$ having the same properties as for the $s<1$ case), as it is predicted by one of our analytical approaches.

In summary, it can be said that the MET switches from polynomial to exponential scaling at $s=1$. This agrees with the drift reversal picture from 
\cite{TCH05,CT08}.
So the (global) Moran process and the (local) local update and Fermi process behave similar, but the impact of the parameters $s$ and $\omega$ is much stronger for the Moran process.

\section{Analytical approaches}
An exact analytical solution for two- or higher dimensional Markov chains is impossible, and a direct computation of the MET of such a three-strategy evolutionary game is not feasible. Although some efforts have been made, 
for example the work of Reichenbach et al. 
\cite{reichenbach}, 
wich have analyzed mean extinction properties in an urn model of a three-strategy game,
employing the usual approach of applying van Kampen's linear noise approximation 
and deriving a Fokker-Planck equation
\cite{van kampen,helbing92,helbing93,TCH05}. 

Adapting this approach to the zero-sum RPS for the local update and the Moran process, we find a predicted MET proportional to $N^2$ and independent of $\omega$, which is in accordance with the simulation data. Unfornately this approach has some shortcomings: 
Although the overall scaling is correct, the numerical value (prefactor) 
of the predicted MET is too great by a factor of $\approx5$, and it is not possible to use this approach for the non zero-sum RPS or in the Fermi process. For that we have developed two 
analytical approaches which do better. We will present both approaches for the local update. Following the same schemes for the other two processes is, although possible, a bit more sophisticated because not all integrals can be done 
analytically here in general.
As we have seen from the numerical investigations, the $s\neq{}1$ payoffs
override differences between the underlying processes, so that it is
warranted to concentrate on one analytically more feasible process in the
remainder.

\subsection{First approach - Expected changes in the distance to the fixed point\label{erster}}

\begin{figure}
\subfigure[]{\epsfig{file=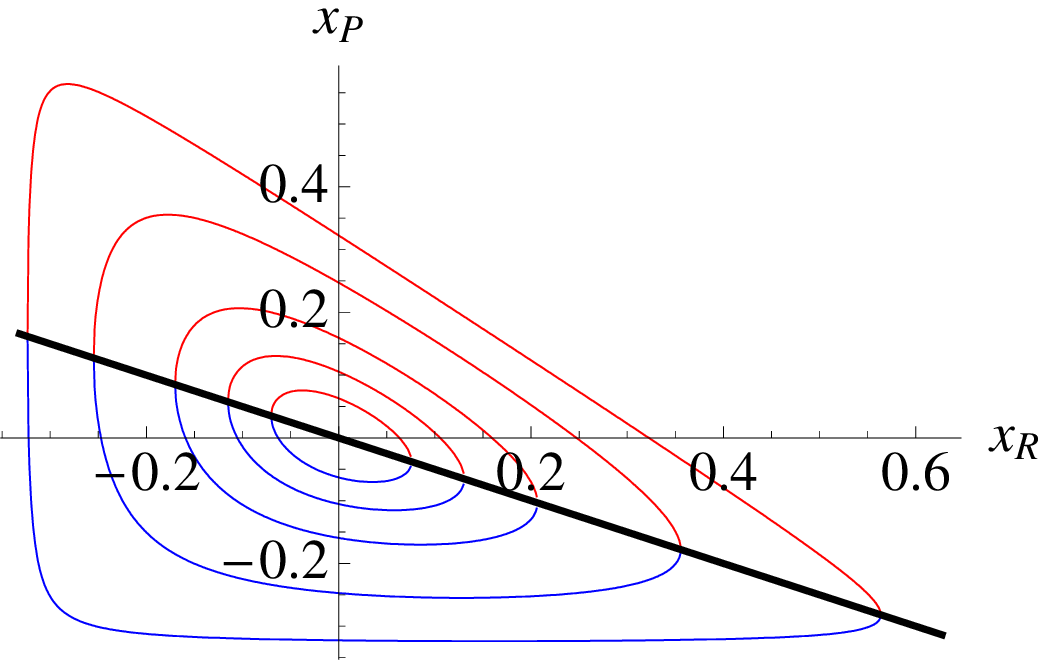,height=50mm}
\label{met_local_einfach}
}
\subfigure[]{\epsfig{file=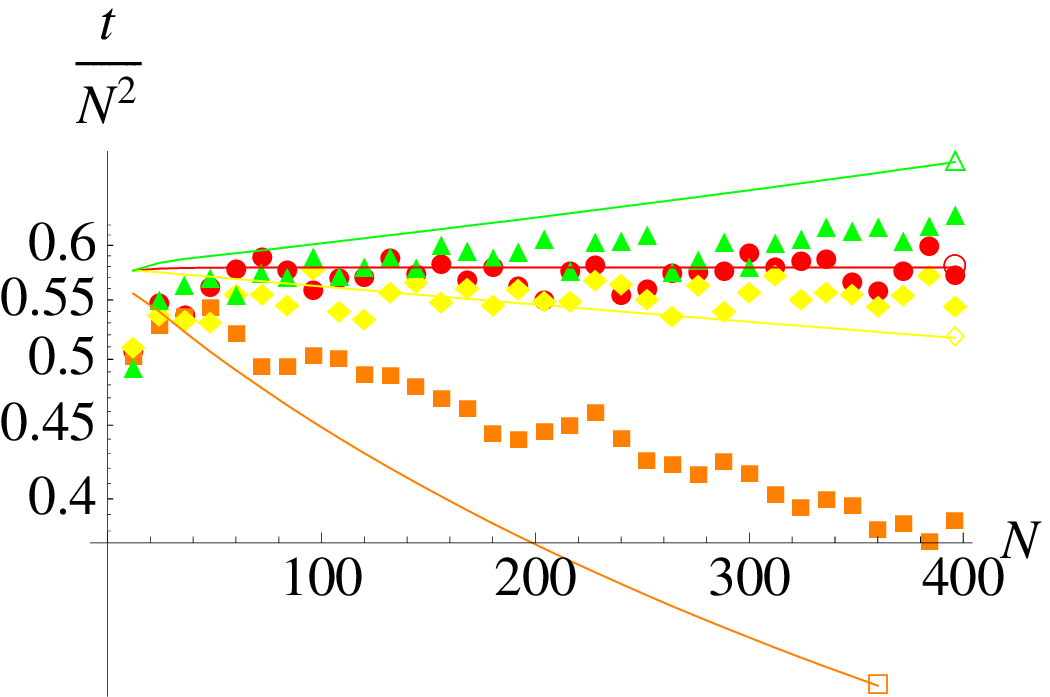,height=50mm}
\label{Ei}
}
\caption{(a) A plot of curves for which $H=const.$ holds, for different values of $H$. 
Different colours on each curve mark the different solution branches $y_u$ und $y_d$. 
The black line connects the values  $x_{min}$ and $x_{max}$, respectively. 
(b) Plots of the simulation data for the MET from the local update (symbols) 
for different pairs of $s$ and $\omega$ depending on $N$ and the corresponding results from the analytical approach presented in section \ref{erster} (lines, with a not filled version of the corresponding symbol at the end). Red circles: $s=1.0$, $\omega=0.5$, orange squares: $s=1.2$, $\omega=0.5$, yellow triangles: $s=1.2$, $\omega=0.05$, green rhombi: $s=0.8$, $\omega=0.05$.}
\end{figure}

To compute -- with help of approximations -- the MET for the general case, we need an appropriate coordinate system. For the standard replicator equation
 and $s=1$,
\begin{equation}
H=-x_R x_P x_S
\end{equation}
is a constant of motion which assumes the value $H=-1/27$ in the fixed point and 
$H=0$ on the edge of the phase space, the simplex $S_3$. 
Here, $x_R$, $x_P$ and  $x_S$ are the frequencies of the strategies $R$, $P$ and $S$, respectively. 
For $s<1$, $H$ is a Lyapunov function of the replicator equations with $\dot H<0$, and so the inner fixed point ist asymptotically stable 
\cite{CT08}. For $s>1$ the fixed point is unstable, and the attractor of the system moves to the edge of the simplex. Via a transformation of the fixed point to the origin of the phase space, and by inserting $x_R+x_P+x_S=1$, which guarantees the conservation of the total density, we find for $H$:
\begin{equation}
H=-(x+1/3)(y+1/3)(1/3-x-y)
\end{equation}
For large absolute values of $H$, the curves with $H=const.$ resemble slightly deformed circles, but for $H\rightarrow 0$ they approach the triangle form of the simplex. In the following we will use $H$ as an observable to measure the effective distance to the fixed point. 
But obviously we need a second coordinate to describe a point in the phase space, so we will use a system with the variables $x$ and $H$ by eliminating the $y$-coordinate. 
This gives us two solutions for $y(x,H)$,
an upper branch above the fixed point, and a corresponding lower branch below. 
We need both because it is not possible to describe all points of the phase space $(x,y)$ by only one solution (see figure \ref{Ei}). So we have
\begin{equation}
y_{u}=\frac{-3x-9x^{2}-\sqrt{4+108H+12x+324Hx-27x^{2}-54x^{3}+81x^{4}}}{6(1+3x)}
\end{equation}
and
\begin{equation}
y_{d}=\frac{-3x-9x^{2}+\sqrt{4+108H+12x+324Hx-27x^{2}-54x^{3}+81x^{4}}}{6(1+3x)}. 
\end{equation}
Here the indices `$u$' and `$d$' stand for `up' and `down', respectively, because by re-inserting of $x$ and $H$ in the first (second) equation one will get only values of $y$ that lie above (under) a separator. In two points, $x_{min}$ and  $x_{max}$, both solutions are identical so that the curve $H=const.$ is closed. 
We can compute these from the condition $y_{u}=y_{d}$ and solving for $x$,

\begin{eqnarray}
x_{min}&=&\frac{1}{3} \left(-\cos \left(\frac{1}{3} \arg \left(-54 H+6 \sqrt{3} \sqrt{H (27
   H+1)}-1\right)\right)\right.\\
&&\left.-\sqrt{3} \sin \left(\frac{1}{3} \arg \left(-54 H+6 \sqrt{3} \sqrt{H
   (27 H+1)}-1\right)\right)+1\right)\nonumber
\end{eqnarray}
and
\begin{eqnarray}
x_{max}&=&\frac{1}{3} \left(-\cos \left(\frac{1}{3} \arg \left(-54 H+6 \sqrt{3} \sqrt{H (27
   H+1)}-1\right)\right)\right.\\
&&\left.+\sqrt{3} \sin \left(\frac{1}{3} \arg \left(-54 H+6 \sqrt{3} \sqrt{H
   (27 H+1)}-1\right)\right)+1\right)\nonumber
\end{eqnarray}
and a third (formal) solution which here is of no use because for typical values of $H$ it produces values outside the simplex. $\arg(z)$ is the argument of the complex number $z$.

Now we compute the expectation value of the change in $H$ for a single time step for the local update process. For time $t$, the system may be in an arbitrary allowed point $(x,H)$. For that we will need the hopping rates in the $(x,H)$ coordinates. For example, for the change $\delta_1=\frac{1}{N},\delta_2=-\frac{1}{N}$ the hopping rate reads after inserting $y=y_u$, 
\begin{equation}
T_u\left(\frac{1}{N},-\frac{1}{N}\right)=-\frac{\left(-9 x^2+3 x+f+2\right)}{216 (3 x+1)} \left(27
   \psi  x^2+9 (\psi -2) x+(2 s+1) f \psi
   -6\right)
\end{equation}
with
\begin{equation}
f=\sqrt{(3 x+1) \left((3 x+1) (2-3 x)^2+108 H\right)}.
\end{equation}
Note that we will use all variables as if they were continous, although in fact they are not, and we use $T(\delta_1,\delta_2)$ as an abbreviation for $T(x_1,x_2,x_3\rightarrow x_1+\delta_1,x_2+\delta_2,x_3+\delta_3)$ due to the conservation of total density (where $\delta_{1}$ und $\delta_{2}$ can take all values of $\{-1/N,0,1/N\}$ as long as they are not identical).

\begin{figure}
\subfigure[]{\epsfig{file=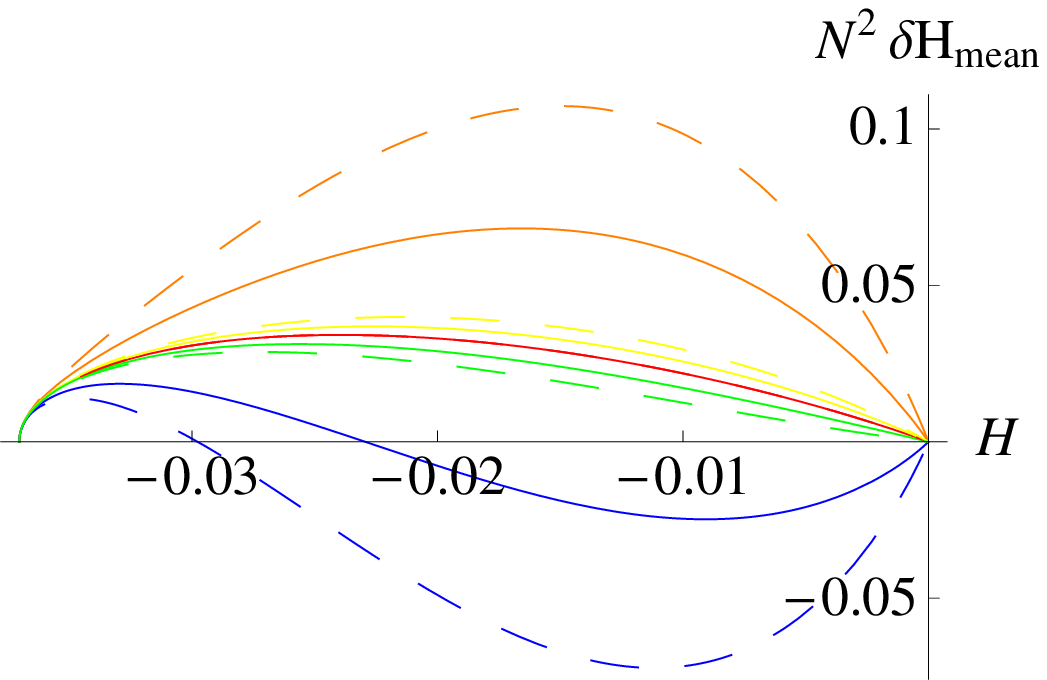,height=50mm}}
\subfigure[]{\epsfig{file=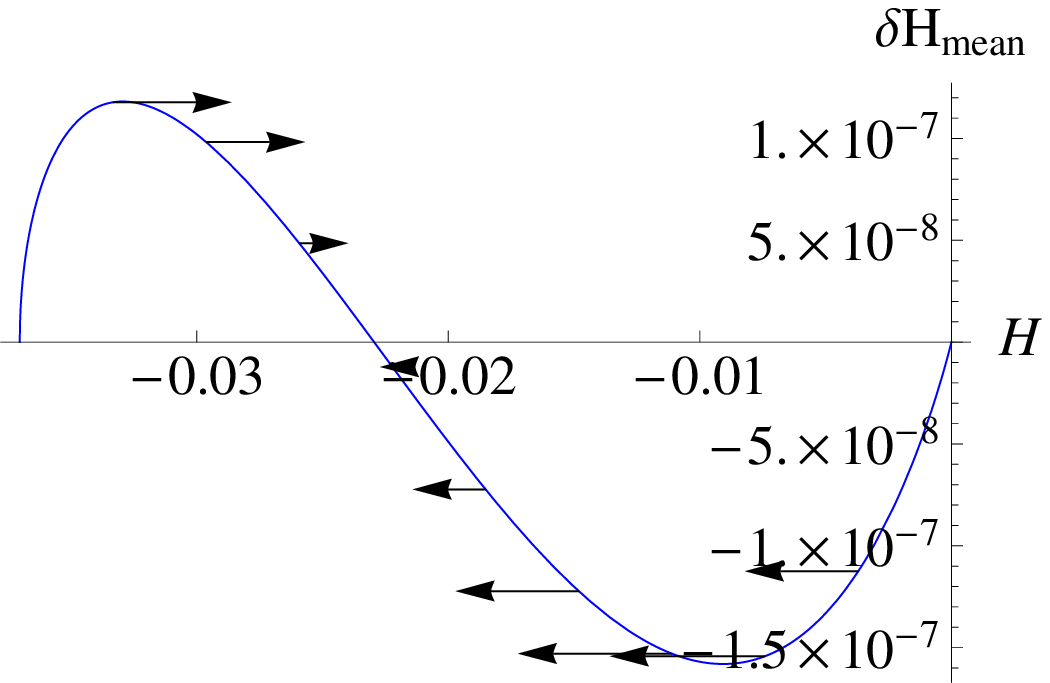,height=50mm}}
\caption{(a) A plot of the expectation values of the changes in $H$, $\delta H(H)$, multiplied with $N^2$, for the local update process. The dashed lines show this quantity for $N=798$, the continous lines for $N=396$. red : $s=1.0$, $\omega=0.5$, orange: $s=1.2$, $\omega=0.5$, yellow: $s=1.2$, $\omega=0.05$, green: $s=0.8, \omega=0.05$, blue: $s=0.8$, $\omega=0.5$. (b) Additional illustration of the meaning of $\delta H(H)$ ($s=0.8$, $\omega=0.50$). If the observable $h$ has a certain value $H_t$ at a time $t$, $H$ is expected to make a step in the direction of the arrow belonging to this specific value of $H$. 
The lengths of the arrows are over-emphasized by a factor of $40000$.}
\label{Woben}
\end{figure}

Further on we will need the change in $H$ that emerges from a change of the $(x,y)$-values in the $(x,H)$-coordinates as well. To derive this, we write $H(x+\delta_{1},y+\delta_{2})$ of the changed values of $x$ and $y$, subduct the initial value $H(x,y)$, and transform the result into the $(x,H)$-coordinates. For $y=y_u$ and the same change as in the example above we get
\begin{equation}
\Delta_u\left(\frac{1}{N},-\frac{1}{N}\right)=\frac{\left(-27 N x^2-9 (N+2) x+N g-6\right)}{36 (3 x
   N+N)^2}  \left(9 x^2-3 x+f-2\right)
\end{equation}
with
\begin{equation}
g=\sqrt{(3 x+1) \left(27 x^3-27 x^2+108 H+4\right)}
\end{equation}
By multiplying all hopping rates $T\left(\delta_{1},\delta_{2}\right)$ with the associated changes in $H$, $\Delta\left(\delta_{1},\delta_{2}\right)$, and summing over all terms, we get a relatively simple equation for the 
expectation value of the change in $H$ in a single time step,
\begin{eqnarray}
\lefteqn{\delta H(x,H)=\sum_{\delta_1,\delta_2}T\left(\delta_{1},\delta_{2}\right)\Delta\left(\delta_{1},\delta_{2}\right)}\\
&&=-\frac{H(9+27x+(1+27H)N(-1+s)\frac{\omega}{\Delta \pi_{max}}+27N(-1+s)x^{3}\frac{\omega}{\Delta \pi_{max}})}{3N^{2}(1+3x)},\nonumber
\end{eqnarray}
with $\delta H_u\left(x,H\right)=\delta H_d\left(x,H\right)=:\delta H\left(x,H\right)$. 
Averaging over $x$ by integrating over $x$ from $x_{min}$ to $x_{max}$ 
and normalizing by the length of whole interval we derive
\begin{eqnarray}
\delta H(H)&=&\frac{\left(6 a^2+(6 b-3) a+6 b^2-3 b+2\right) N (s-1) \psi +18}{6N^2}\nonumber\\
&&-\frac{3 H (s-1) \psi  \log \left(\frac{3 b+1}{3 a+1}\right)}{(a-b)N}\label{delta_H_von_H}
\end{eqnarray}
with $a=x_{min}$, $b=x_{max}$. The simplest way to approximate the mean extinction time for a system starting in the fixed point of the replicator equations is then to define a map
\begin{equation}
H_{i+1}=H_i+\delta H(H_i)
\end{equation}
and to iterate it until $H_{\langle t\rangle}\approx 0$ is reached. The MET is then simply the number of necessary steps. 

As one can see in the right hand side of Figure \ref{met_local_einfach}, 
this result is in good agreement with the simulation data, for the 
dependence of $s$ and $\omega$, as well as for the numerical values. For $s=1$ the MET is predicted to be proportional to $N^2$ and independent of $\omega$ because $\delta H(H)$ is independent of $\omega$ for $s=1$. It is in good agreement with the simulations that the MET for $s\neq 1$ and weak selection differs only slightly from the MET for $s=1$, while for strong selection and $s>1$ it is clearly smaller than for $s=1$.

Unfornately, for $s<1$ this approach works only for small selection and not to great $N$, because for $s<1$ non-positive expectation values of the changes in $H$ are possible, so if we start the mentioned iterated map in the fixed point and repeat the iteration again and again, we would arrive at a value of $H$ where the expected change is $0$ at some time, and the predicted MET would be infinite. The part of the interval where the expectation value is negative is bigger for 
larger $\omega$ 
than for smaller $\omega$, and it grows with increasing $N$, 
so in this cases this approach is not sufficient.

\subsection{Second approach - Fokker-Planck equation\label{zweiter}}

For the local update process, a Fokker-Planck equation (FPE) for the time evolution of the 
probability density is given by (summation over double indices)
\begin{equation}
\partial_{t} P(\vec{x},t)=-\partial_{i}[\alpha_{i}(\vec{x})P(\vec{x},t)]+\frac{1}{2}\partial_{ij}[B_{ij}(\vec{x})P(\vec{x},t)],
\end{equation}
where the coefficients are
\begin{equation}
\alpha_{i}=\sum_{\delta \vec{x}}\delta x_{i}T(\vec{x}\rightarrow \vec{x}+\delta \vec{x})\end{equation}
and
\begin{equation}B_{ij}(\vec{x})=\sum_{\delta \vec{x}}\delta x_{i}\delta x_{j}T(\vec{x}\rightarrow \vec{x}+\delta \vec{x}),
\end{equation}
which gives the following results 
\begin{eqnarray}
\alpha_{R}&=&-\frac{(2x_{P}+x_{R})(1+3x_{R})\psi}{3N}\\
\alpha_{P}&=&\frac{(1+3x_{P})(x_{P}+2x_{R})\psi}{3N}\\
B_{RR}&=&\frac{2+3x_{R}-9x_{R}^{2}}{9N^{2}}\\
B_{RP}&=&B_{PR}=-\frac{(1+3x_{P})(1+3x_{R})}{N^{2}}\\
B_{PP}&=&\frac{2+3x_{P}-9x_{P}^{2}}{9N^{2}}.
\end{eqnarray}
Transforming this into polar coordinates $x_R=r\cos\phi$, $x_P=r\sin\phi$, we find for the coefficients of the FPE
\begin{flalign}
\alpha_R&=-\frac{r \psi  (3 r \cos (\phi )+1) (2 \cos (\phi )+2 (s+1) \sin (\phi )+r (s-1) (\sin (2 \phi )+2))}{6 N}&\\
\alpha_P&=-\frac{r \psi  (3 r \sin (\phi )+1) (-2 (s+1) \cos (\phi )-2 s \sin (\phi )+r (s-1) (\sin (2 \phi )+2))}{6 N}&\\
B_{RR}&=\frac{-9 r^2 \cos ^2(\phi )+3 r \cos (\phi )+2}{9 N^2}&\\
B_{RP}=B_{PR}&=-\frac{(3 r \cos (\phi )+1) (3 r \sin (\phi )+1)}{9 N^2}&\\
B_{PP}&=\frac{-9 r^2 \sin ^2(\phi )+3 r \sin (\phi )+2}{9 N^2},&
\end{flalign}
where $\psi=\omega/\Delta\pi_{max}$, and hence for the FPE itself
\small
\begin{flalign}
P^{(0,0,1)}&=\frac{6 N (s-1) \psi  \sin (2 \phi ) r^2+N \left(12
   r^2-1\right) (s-1) \psi -9}{3 N^2}P^{(0,0,0)}&\\
&+\left(\frac{r \psi  (-3 (s+2) \cos (\phi )+(7 s+2) \cos (3
   \phi )-2 (4 s+(2 s+7) \cos (2 \phi )+5) \sin (\phi
   ))}{12 N}\right.\nonumber&\\
&\left.-\frac{(s+1) \psi  (\sin (2 \phi )+2)}{6 N}+\frac{\cos (\phi )+3 \cos (3 \phi )-\sin (\phi )+3
   \sin (3 \phi )}{12 N^2 r}+\frac{\cos (2 \phi )}{9 N^2 r^2}\right)P^{(0,1,0)}\nonumber&\\
&+\left(\frac{r^2 \psi  ((2 s+1) \cos (\phi )+(2 s+7) \cos (3 \phi )-(s+2)
   \sin (\phi )+(7 s+2) \sin (3 \phi ))}{12 N}\right.\nonumber&\\
&\left.+\frac{r^3 (s-1) \psi  (\cos (\phi ) \sin (\phi )+1)}{N}+\frac{r (-N (s-1) \psi +N (s+1) \cos (2 \phi ) \psi -18)}{6 N^2}\right.\nonumber&\\
&\left.+\frac{\cos (\phi )-\cos (3 \phi )+\sin (\phi )+\sin (3 \phi )}{8
   N^2}+\frac{\cos (\phi ) \sin (\phi )+1}{9 N^2 r} \right)P^{(1,0,0)}\nonumber&\\
&-\frac{(\cos (\phi )-\sin (\phi )) (9 \cos (\phi )
  \sin (\phi ) r+3 r+\cos (\phi )+\sin (\phi ))}{9
   N^2 r}P^{(1,1,0)}\nonumber&\\
&+\frac{9 r \cos (\phi )-9 r \cos (3 \phi )+9 r \sin
   (\phi )+4 \sin (2 \phi )+9 r \sin (3 \phi )+8}{72
   N^2 r^2}P^{(0,2,0)}\nonumber&\\
&+\frac{-36 r^2+3 \cos (\phi ) r+9 \cos (3 \phi ) r+3
   \sin (\phi ) r-9 \sin (3 \phi ) r-4 \sin (2 \phi
   )+8}{72 N^2}P^{(2,0,0)},\nonumber&
\end{flalign}
\normalsize
with $P^{(\lambda,\mu,\nu)}:=\frac{\partial^\lambda}{\partial r^\lambda}\frac{\partial^\mu}{\partial\phi^\mu}\frac{\partial^\nu}{\partial t^\nu}P(r,\phi,t)$.

If we now assume the probability density to be independent of the angle $\phi$ (which is a good approximation at least for small $r$), all terms which contain derivatives with respect to $\phi$ will drop out. By averaging over $\phi$ afterwards by integrating from $0$ to $2\pi$ over $\phi$ and dividing the result by $2\pi$ we get a probability density which is only independent of $r$ and $t$. 
Hence the FPE reads
\begin{eqnarray}
\lefteqn{\partial_t P(r,t)=\frac{r \left(N \left(12 r^2-1\right) (s-1) \psi
   -9\right)}{3 N^2 \sqrt{r^2}}P(r,t)}\\
&&+\frac{3 \left(N \left(6 r^2-1\right) (s-1) \psi
   -18\right) r^2+2}{18 N^2 \sqrt{r^2}}\partial_r P(r,t)\nonumber\\
&&+\frac{r \left(2-9 r^2\right)}{18 N^2 \sqrt{r^2}}\partial_{r}^2P(r,t)\nonumber\\
&&=:a(r) P(r,t)-v(r)\partial_r P(r,t)+D(r)\partial_r^2 P(r,t).
\label{1D-Fokker-Planck}
\end{eqnarray}
Let us now approximate the diffusion by its value at $r=0$, this gives us
\begin{equation}
D=\frac{1}{9N^2}.
\end{equation}
$r$ can theoretically vary between $0$ and $2/3$ for some values of $\phi$, 
but most extinction events will happen around those point of the edge of the simplex that are closest to the inner fixed point. These points have the distance $r\gtrsim \frac{1}{3}$ from the inner FP. Because of that we search for the solution of the one-dimensional random walk descripted by eq. \ref{1D-Fokker-Planck} in the interval $[0,L]=[0,\frac{1}{3}]$. Next, let us approximate the convection velocity $v(r)$ by its value for mediate $r$, e.g. $r=\frac{1}{6}=\frac{L}{2}$, and we get
\begin{equation}
v=\left\{\begin{array}{cc}
\frac{-0.00462963 N \psi -0.166667}{N^2},&s=0.8\\&\\-\frac{1}{6 N^2},&s=1.0\\&\\\frac{0.00462963 N \psi -0.166667}{N^2},&s=1.2
\end{array}
\right.
\end{equation}
\begin{figure}
\epsfig{file=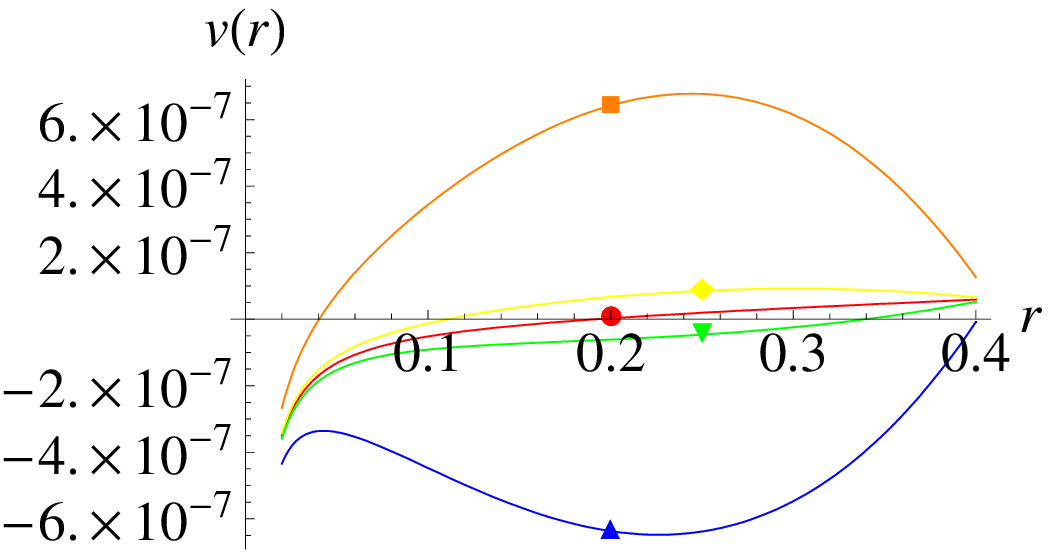,height=50mm}
\epsfig{file=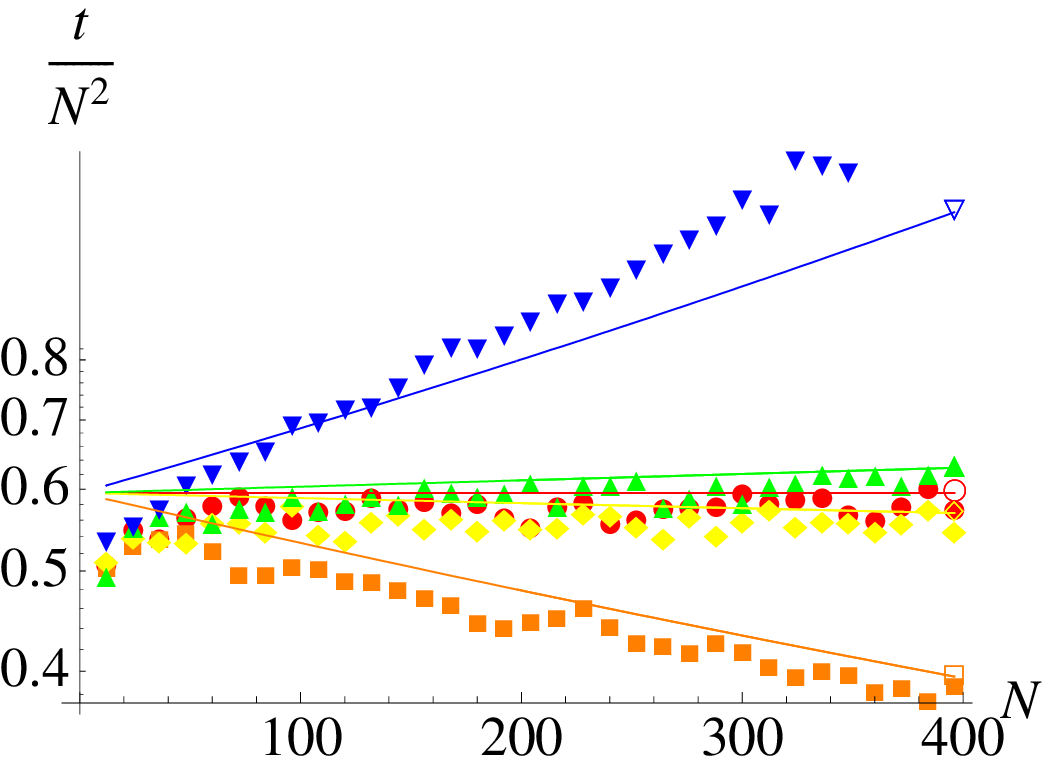,height=50mm}
\caption{Left: A plot of the convection velocity $v(r)$ for the local update with $N=3960$. As one can see, $v(r)$ is $\approx 0$ for $s=1$. This corresponds to the neutral stability of the fixed point in the limit $N\rightarrow \infty$ for $s=1$. For $s<1$ $v(r)$ is negative, corresponding to the asymptotical stable FP, and for $s>1$ $v(r)$  is positive corresponding to the 
unstable FP. For $r\rightarrow 0$ $v(r)$ diverges for all parameter values to $-\infty$, which results from the simplifications in combination with the existence of a FP at $r=0$.  Right: Semilogarithmic plot of simulation data of the MET in the local update process, divided by $N^2$ (symbols) for several pairs of $s$ and $\omega$, and the corresponding values predicted by the approach in section \ref{zweiter} (lines, with a not filled version of the corresponding symbol at the end). Red circles: $s=1.0$, $\omega=0.5$, orange squares: $s=1.2$, $\omega=0.5$, yellow rhombi: $s=1.2$, $\omega=0.05$, green upward triangles: $s=0.8$, $\omega=0.05$, blue downward triangles: $s=0.8$, $\omega=0.5$}
\label{v}\label{MET_theo}
\end{figure}
(We cannot approximate it by its value at $r=0$ as we have done for the diffusion because $v(0)$ is $-\infty$ corresponding to the FP.) For a last simplification, we neglect the term that includes the probability density itself. 
Hence, the FPE reads
\begin{equation}
\partial_t P(r,t)=-v\partial_r P(r,t)+D\partial_r^2 P(r,t).
\end{equation}
The computation of the MET for such an equation with one reflecting border at $r=0$ and one absorbing border at $r=L$ is a standard problem, for example, see 
\cite{redner}. One can find the MET by introducing a time integrated probability density,
\begin{equation}
P_0(r)=\int_0^{\infty}P(r,t)dt.
\end{equation}
With this, the time-integrated FPE reads:
\begin{equation}
v \frac{\partial P_0(x)}{\partial r}=D \frac{\partial^2 P_0(x)}{\partial x^2}
\end{equation}
with the boundary conditions $P_0(r)=0$ at $r=L$ and $J(r)=\int_0^{\infty}j(r,t)dt=\int_0^{\infty}v P(r,t)-D\partial_r P(r,t)dt=1$ at $r=0$. There, $J(r)$ is the time-integrated probability current. The constraint $J(0)=1$ corresponds to the injection of an unit probability current at $t=0$, this means a start of the system in the origin. The solution of the differential equation is given by
\begin{equation}
P_0(r)=\frac{1}{v}\left(1-e^{v(x-L)/D}\right)
\end{equation}
Then the MET is simply
\begin{equation}
\langle t\rangle=\int_0^L P_0(r)dr=\frac{L}{v}-\frac{D}{v^2}(1-e^{-vL/D}),
\end{equation}
what leads to
\begin{equation}
\langle t\rangle=
\left\{\begin{array}{cc}
\frac{N^2 \left(-0.00154321 N \psi +0.183191
   e^{0.0138889 N \psi }-0.166667\right)}{(0.00462963
   N \psi +0.166667)^2},&s=0.8\\&\\
0.594885 N^2,&s=0\\&\\
\frac{e^{-0.0138889 N \psi } N^2 \left(e^{0.0138889 N
   \psi } (0.00154321 N \psi
   -0.166667)+0.183191\right)}{(0.166667-0.00462963 N
   \psi )^2},&s=1.2
\end{array}
\right.
\end{equation}
or in general
\begin{eqnarray}
\langle t\rangle=\frac{72 e^{-\frac{5}{72} N (s-1) \psi -\frac{1}{2}}
   N^2 \left(e^{\frac{5}{72} N (s-1) \psi
   +\frac{1}{2}} (5 N (s-1) \psi -108)+72
   e\right)}{(36-5 N (s-1) \psi )^2}.
\end{eqnarray}
The proportional dependence of the MET on $N$ and the other parameters is very good, as one can see in Fig. \ref{MET_theo}. This approach predicts the MET to be proportional to $N^2$ and independent of the selection strength $\omega$ for $s=1$, while for $s<1$ it forecasts the MET to be substantially proportional to $N^2 e^{\kappa_1\psi N}$, and for $s>1$ in good approximation proportional to $c_1 N^2+c_2 e^{\kappa_2\psi N}$.
\section{Summary}
Evolutionary games with three 
cyclically dominating strategies
can
stabilize the coexistence of strategies 
superior
compared to games with 
only two strategies which include a dominating strategy.
In finite populations, it is still likely that one of these strategies 
goes extinct, though on average it takes a time which is considerably longer than in two strategy games. We have analyzed the mean 
times to the extinction of one of the three strategies numerically in the RPS-game for three different evolutionary processes, 
the local update, the frequency-dependent 
Moran process and the Fermi process. These mean extinction times (MET) have the same 
fundamental dependencies of the number of agents $N$, the parameter $s$ and the selection strength $\omega$.

For the zero-sum RPS game ($s=1$), the MET is proportional to $N^2$ and independent of $\omega$ 
for all three processes;
even the proportionality constants are the same. 
For $s<1$, we find a stabilization of biodiversity, as the MET grows exponentially with $N$. 
In contrast to that, for $s>1$, there is a destabilization of biodiversity, and the MET now grows smaller than in the 
zero-sum RPS (where it grows $\propto N^2$), namely, a part of the proportionality factor of the 
zero-sum RPS now decays exponentially with $N$. 
In both non zero-sum cases, the coefficient that is multiplied with the $N$ 
in the exponent, grows with both $\omega$ and $(s-1)$.

Solving such two-dimensional Markov chains is not possible, but we have developed two analytical approaches for the approximation of the dependencies of the MET of $N$ and the parameters $s$ and $\omega$. 
The first approach can be used for $s\geq 1$ only, 
but serves for all three processes. 
In contrast, the second approach gives a good approximation for all parameter values, but it can be used completely analytical only for the local update process, as for the other two processes not all integrals can be done analytical.

In summary, the fixation time -- as a long-time property of the process --
shows a crossover from exponential to polynomial scaling with the population
size, being fully consistent with the critical population size
deived from the ``drift reversal'' picture which is based on the
short-time dependence 
of the average drift in phase space - 
repelling versus attracting towards the fixed point.
Thus both approaches to describe stochastic stability in finite
populations here lead to the same conclusions: 
coexistence is preserved (stabilized) for a positive-sum cyclic game,
whereas negative-sum games as well as small populations destabilize
coexistence.

\end{document}